\documentclass[aps,prd,showpacs,preprintnumbers,nofootinbib,twocolumn]{revtex4}

\usepackage{bm}
\usepackage{latexsym} 
\usepackage{amsfonts,amssymb}
\usepackage{graphicx,epsfig}
\usepackage{psfrag}

\def\beq{\begin{equation}}
\def\eeq{\end{equation}}
\def\br{\begin{eqnarray}}
\def\er{\end{eqnarray}}

\def\l{\left}
\def\r{\right}

\def\anp{{\em  \/Ann.\ Phys.~}}

\def\ijmpd{{\em \/Int.\ Journ.\ Mod.\ Phys.\ D~}}
\def\mpla{{\em \/Mod.\ Phys.\ Let.\ A~}}

\def\pla{{\em  \/Phys.\ Lett.\ A~}}
\def\plb{{\em  \/Phys.\ Lett.\ B~}}
\def\prd{{\em  \/Phys.\ Rev.\ D~}}
\def\prl{{\em  \/Phys.\ Rev.\ Lett.~}}

\begin{document}
\preprint{gr-qc/0203060}

\title[Short Title]{Is there an imprint of Planck scale physics on 
inflationary cosmology?}

\author{S.~Shankaranarayanan\footnote{Present Address: DCTD, University 
of Azores, 9500 Ponta Delgada, Portugal. Email: shanki@notes.uac.pt}}
\affiliation{IUCAA, Post Bag 4, Ganeshkhind, Pune 411 007, INDIA.}

\begin{abstract}
We study the effects of the trans-Planckian dispersion relation on the
spectrum of the primordial density perturbations during inflation. In
contrast to the earlier analyses, we do not assume any specific form
of the dispersion relation and allow the initial state of the field to
be arbitrary. We obtain the spectrum of vacuum fluctuations of the
quantum field by considering a scalar field satisfying the linear wave
equation with higher spatial derivative terms propagating in the de
Sitter space-time. We show that the power spectrum does not {\it
strongly} depend on the dispersion relation and that the form of the
dispersion relation does not play a significant role in obtaining the
corrections to the scale invariant spectrum.  We also show that the
signatures of the deviations from the flat scale-invariant spectrum
from the CMBR observations due to quantum gravitational effects {\it
cannot} be differentiated from the standard inflationary scenario with an
{\it arbitrary} initial state.
\end{abstract}

\pacs{98.80.Cq, 98.70.Vc}

\maketitle

\section{Introduction}

The present upper limits on the temperature fluctuations of the
microwave background radiation, corresponding to the density
perturbations at the time of recombination, are less than $1$ part in
$10^5$ on scales larger than the Hubble radius. On these scales, the
observations of the CMBR gives the power spectral index $n$ to be $1.1
\ \pm \ 0.1$. These observations provide an upper bound on the
amplitude of density fluctuations at that epoch. Inflationary paradigm
provides a physical mechanism for the generation of seed density
perturbations (see, for example, Ref. \cite{linde}).  Many models of
inflation require a period much longer than 70 e-foldings to solve the
horizon and flatness problems of standard cosmology. In order to
obtain the observed density inhomogeneities on the galactic scales,
these models of inflation require the modes of vacuum fluctuations at
the initial epoch of inflation to be smaller than Planck length $L_P
\equiv (G\hbar/c^3)^{1/2}$ \cite{paddyold}.

It is well known that the space-time structure at Planck scales is
affected by the quantum gravitational effects and it is
generally believed that $L_P$ acts as a physical cutoff for space-time
intervals. The existence of a fundamental length implies that
processes involving energies higher than Planck energies will be
suppressed, and the ultraviolet behavior of the theory will be
improved. All existing models of quantum gravity provide a mechanism
for good ultra-violet behavior, essentially through the existence of
the fundamental length scale. Even though the models of quantum
gravity have some success, none of these have given a complete theory
that works at the Planck scales.

If the ultimate theory of quantum gravity has a fundamental length
scale, then the low-energy effective quantum field theories should
have an imprint of the Planck scale. However, the standard formulation
of quantum field theory does not take into account the existence of
any fundamental length in the space-time. Attempts have been made to
incorporate quantum gravitational effects with the help of two
different approaches: (i) by introducing the fundamental length scale
in a Lorentz-invariant manner in the Feynman Green function and (ii)
by modifying the dispersion relation of the linear field equation.

In this work, our focus is on approach (ii). (For method (i) and its
implications in conventional quantum field theory, see
Ref. \cite{paddy1,paddy2}.) In the context of inflationary cosmology,
trans-Planckian effects have gained interest after the detailed work
of Martin and Brandenberger \cite{branden,branden1} who introduced
modified dispersion relations in an {\it adhoc} manner. The authors
dealt with high-frequency dispersion relations, used earlier by Unruh
\cite{unruh} and Corley/Jacobson \cite{corley} for the analysis of
deviations from the thermality of the Hawking spectrum, to the
inflaton field and obtained the spectrum of the vacuum fluctuations
after it crosses the Hubble radius (during the inflation). They found
that while some hypothesized dispersion relation(s) to quantum field
theory yield no change in the power spectrum of the fluctuations, some
indicate the cosmological sensitivity to Planck scale
physics. However, there have been arguments that the spectrum of
vacuum fluctuations should not be sensitive to the Planck scale
physics \cite{staro,paren}.

All previous analyses of trans-Planckian effects on inflationary
perturbations \cite{branden}-\cite{other_pap} have concentrated on
specific form of the dispersion relations for a particular choice of
initial state of the field. [Two sets of initial conditions which are
used in the literature are Minimum energy and Instantaneous Minkowski
vacuum.] In our analysis, we do not assume any specific form of the
dispersion relation and also allow initial state of the field to be
arbitrary. The only condition on the dispersion relation being that
the non-linear dispersion relations become linear for $\lambda >>
L_P$. From the earlier analyses, we would expect that the spectrum of
fluctuations will strongly depend on the nature of the dispersion
relation and the initial state of the field. 

Using this strategy, we address the following questions: Is there an
imprint of Planck scale physics on inflationary cosmology? {\it and}
Can the deviations from the flat scale-invariant spectrum due to
non-linear dispersion relations be differentiated from that of deviations
from the flat spectrum in the standard inflationary scenario? We try
to provide an answer by considering a quantum field satisfying the
linear wave equation with higher spatial derivative terms propagating
in the de Sitter space-time. We perform a general analysis for a class
of dispersion relations whose field modes can be expanded as WKB modes
over a range of initial states (conditions) for the quantum field. [In
this paper we are primarily interested in frequency evolution of modes
and we ignore the actual creation of the modes. There are claims in
the literature that the energy density of fluctuations of the inflaton
field is necessarily high and its back-reaction on the metric cannot
be ignored
\cite{tanaka,staro} (see also, Ref. \cite{tpsingh}).] 

We show that the signatures of the deviations from the flat
scale-invariant spectrum due to non-linear dispersion relations {\it
cannot} be differentiated from the standard inflationary scenario with an
{\it arbitrary} initial state. We supplement this claim by studying four
dispersion relations:
\br
\label{eq:paddy_dis}
\Omega^2& =& \frac{E_P^2}{2\pi} 
\log \left(1 + \frac{2\pi k^2}{E_P^2}\right) \\
& & {\rm (Padmanabhan's~ dispersion~ relation)} \nonumber \\
\label{eq:kappa_dis}
\Omega^2& =& \kappa^2 \log^2(1 + k/\kappa), \\
& &  (\kappa-{\rm Poincare~dispersion~relation}) \nonumber \\
\label{eq:unruh_dis}
\Omega^2 & = & k_{\rm Pl}^2 \tanh[(k/k_{\rm Pl})^n]^{(2/n)}, \\
& &  ({\rm Unruh's~dispersion~relation}) \nonumber \\
\label{eq:ahlu_dis}
\Omega^2 & = & k_{\rm Pl}^2 \tan^{-2}[k/k_{\rm Pl}], \\
& &  ({\rm Ahluwalia's~dispersion~relation}) \nonumber
\er
where $E_P$ is the Planck energy and $\kappa, k_{\rm Pl}$ is the
inverse Planck length ({\it or} of the order of Planck scale). The
dispersion relation (\ref{eq:paddy_dis}) reproduces the exact density
of states required for the black-hole by treating black-hole as a
one-particle state of a non-local field theory \cite{paddy98},
relation (\ref{eq:kappa_dis}) is motivated by $\kappa$-Poincare
algebra
\cite{gilkman} while the last relation is the gravitationally modified 
expression for the wave-particle duality \cite{aluwalia}. 

The organization of the paper is as follows: In Sec. (II), we
calculate the power spectrum of the vacuum fluctuations for an
arbitrary initial condition in (i) standard inflationary scenario (ii)
dispersive field theory models, and interpret the results. Finally, in
Sec. (III) we summarize the results.


\section{Power spectrum of quantum fluctuations}

In this section, we calculate the power spectrum of fluctuations for
an arbitrary initial state of a quantum field $\Phi$, in standard
inflationary scenario and in dispersive field theory models,
propagating in the spatially flat FRW background (specifically de
Sitter). We show that using the observations of CMBR the signatures of
the deviations from the flat scale-invariant spectrum due to
non-linear dispersion relations {\it cannot} be differentiated from the
standard inflationary scenario with an {\it arbitrary} initial state.
\begin{table*}[!hbt]
\caption{The dominant contribution and solution to the 
Eq. (\ref{eq:nonlin_scalar}) in three different regimes. In this
$\eta_i$ is the epoch at which inflation starts, $\eta_{Pl}$ is the
epoch at which the non-linear dispersion relation can be approximated
to be linear and $\eta_H$ is the epoch at which the perturbation
leaves the Hubble radius.}
\label{ta:sum}
\begin{ruledtabular}
\begin{tabular}{cccccccc}
Regime&Epoch & Dominant contribution of $\varpi^{\rm mod}_k(\eta)$ &
Solution \\ \hline \\

I&$\eta_{Pl}>> \eta \ge \eta_i$ & 
$ W^I_k(\eta) = a(\eta)\Omega[k_{\rm ph}(\eta)]$ & 
$\chi^I_k \sim\l[2 W^I_k(\eta)\r]^{-1/2} \exp\l(\pm i \int W^I_k(\eta) d\eta\r)$ \\
&($ \lambda_{\rm ph} << H^{-1}$) & (non-linear dispersion) & 
(WKB approximate solution) \\
& & & & \\
& & & & \\
II&$\eta_H >> \eta \ge \eta_{Pl}$ & 
$W^{II}_k(\eta) = k $ & $\chi^{II}_k \sim \exp(\pm i k \eta)$ \\
&($ \lambda_{\rm ph} << H^{-1}$) & (linear dispersion) & (Plane 
wave solution)\\
& & & & \\
& & & & \\
III & $\eta \ge \eta_H$ & 
$[W^{III}_k(\eta)]^2 = - \l[a''(\eta)/a(\eta)\r]$ & 
$\chi^{III}_k \sim C_k a(\eta)$ \\
& ($ \lambda_{\rm ph} >> H^{-1}$) & & (frozen classical perturbation) 
\end{tabular}
\end{ruledtabular}
\end{table*}
\subsection{Standard inflation}
\label{subsec:standard}
Let us consider scalar fields propagating in the flat FRW metric of the form
\beq
ds^2= dt^2 - a^2(t) d{\bm{x}}^2 = a^2(\eta)\left[d\eta^2- d{\bm{x}}^2
\right],
\label{eq:frw} 
\eeq
where ${\bm x}$ is the 3-space, $t$ is the comoving time, $\eta$ is
the conformal time and $t = \int a(\eta) d\eta$. For flat de Sitter
$a(\eta) = 1/( H \eta)$ [$a(t) = \exp(H t)$] and $H$ is the Hubble
constant during inflation. [Typical values of the inflationary
expansion are $ E \approx 10^{15} GeV$, 
$H \approx 5 \times 10^{24}\, cm^{-1}$, start of the inflation 
$\eta_i \approx - 10^{-36} \, sec$ ($t_i \approx 10^{-36} \, sec$) 
and the end of inflation $t_f \approx 70 H^{-1}$.] The free massless, 
minimally coupled scalar field satisfies the Klein-Gordon equation,
\beq 
\Box\Phi \equiv \frac{d^2\Phi}{d\eta^2} + 2 \frac1{a^3(\eta)}
 \frac{da(\eta)}{d\eta} - \nabla^2\Phi = 0.
\label{eq:sca_eom}
\eeq
The symmetry of the Robertson-Walker metric allows for separating
variables in Eq. (\ref{eq:sca_eom}) and the scalar field can be 
decomposed as
\beq
\Phi({\bm x}, \eta) = \frac{1}{(2\pi)^{3/2}} \int_k \l[a_k u_k({\bm x}, \eta) 
+ a_k^\dagger u_k^*({\bm x}, \eta)\r] d{\bm k}, 
\label{eq:four_exp}
\eeq
where 
\beq
u_k({\bm x}, \eta) = a(\eta)^{-1} \chi_k(\eta) \exp(i {\bm k}.{\bm x}).
\eeq
[${\bm k}$ is the comoving wave vector and $k^2 = |{\bm k}|^2$.]
Note that the creation and annihilation operators $a_k^{\dagger}, a_k$
obey the usual commutation relations. The time-dependent part of the
mode function satisfies the oscillatory equation
\beq
\frac{d^2\chi_k}{d\eta^2} + \varpi_k^2(\eta) \chi_k(\eta) = 0, 
\label{eq:lin_scalar}
\eeq
where $\varpi^2_k(\eta) = k^2 - [a''(\eta)/a(\eta)]$. [The solution to
the above equation reduces to the usual harmonic oscillator solution,
$\chi_k \sim \exp(i\omega\eta)$, with a linear dispersion relation,
$\omega^2 = k^2_{\rm ph} = k^2/a^2(\eta)$, on scales much smaller than
the Hubble length. $k_{\rm ph}$ is the physical momentum of the
perturbation.] A complete set of mode solutions to
Eq. (\ref{eq:lin_scalar}) is obtained by imposing initial conditions
$\chi_k(\eta_i)$, $\chi'_k(\eta_i)$ on a Cauchy surface $\eta =\eta_i$
corresponding to a homogeneous initial state.

The solution to the above equation in the short-wavelength limit
($\lambda_{\rm ph} << H^{-1}$) corresponds to the Minkowski modes,
i. e.,
\beq
\chi^I = \frac{a_k}{\sqrt{2k}}  \exp(i k \eta) + 
\frac{b_k}{\sqrt{2k}} \exp(-i k \eta)
\label{eq:short_mod}
\eeq
where, $a_k$ and $b_k$ are determined by the choice of the initial
conditions. Orthonormality of the modes gives,
\beq
|a_k|^2 - |b_k|^2 = 1
\label{eq:akbk_cond}
\eeq
Thus, $a_k$ and $b_k$ can be any {\it arbitrary} functions of $k$
satisfying the above condition. In the long wave-wavelength limit
($\lambda_{\rm ph} >> H^{-1}$), the modes are frozen and the solution
is given by
\beq
\chi^{II} =  C_k a(\eta) = \frac{C_k}{H \eta}.
\label{eq:long_wav}
\eeq
Demanding that $\chi^I$ and $\chi^{II}$ are continuous at $\eta_H$,
$C_k$ can be determined in-terms of $a_k$ and $b_k$. [The junction
condition at $\eta_H$ between the two modes correspond to the relation
$\{2 \pi a(\eta)\}/k = H^{-1}$, i. e., $\eta_H = 2\pi/k$.]  Thus, for
the $k$ mode leaving the Hubble radius at $\eta_H$, we get
\beq
k^3 |C_k|^2 = \frac{H^2}{2} + H^2 \l[|b_k|^2 + 
|b_k| (|b_k|^2 + 1)^{1/2} \r], 
\label{eq:C_K_lin}
\eeq
where we have used condition the (\ref{eq:akbk_cond}) to write
$|a_k|^2$ in-terms of $|b_k|^2$.  The above result shows that the
spectrum of fluctuations has a scale-invariant term plus a non
scale-invariant term which strongly depends on the choice of the
initial state ($b_k$). If we assume that on cosmologically relevant
scales {\it the inflaton field is in the vacuum state, corresponding}
{\it to no inflaton particles} corresponding to the initial condition,
\beq
\chi(\eta_i) = \frac{1}{\sqrt{2 k}}; \qquad \qquad
\chi'(\eta_i) = \pm i \sqrt{\frac{k}{2}},
\eeq
we get $a_k = 1, b_k = 0$; which results in a flat scale-invariant
spectrum.

\subsection{Dispersive field theory model} 
\label{subsec:dispersive}

As mentioned in the introduction, in many models of inflation, the
physical length of the perturbations is smaller than the Planck length
at the initial epoch implying we need to consider quantum
gravitational corrections. One way of introducing quantum
gravitational corrections is to modify the linear dispersion relation
{\it in an ad-hoc manner} by breaking the Lorentz invariance i. e., by
rewriting $\omega^2 = \Omega^2[k_{\rm ph}(\eta)]$. Replacing $k^2$ in
Eq. (\ref{eq:lin_scalar}) as
\beq
{k^2_{\rm mod}}  = a^2(\eta) \Omega^2[k_{\rm ph}(\eta)] = a^2(\eta) 
\Omega^2\l(\frac{k}{a(\eta)}\r),
\label{eq:mod_disp}
\eeq
we get,
\beq
\frac{d^2\chi_k}{d\eta^2} + {\varpi^{\rm mod}}_k^2(\eta) \chi_k = 0,
\label{eq:nonlin_scalar}
\eeq
where, 
\beq 
{\varpi^{\rm mod}}_k^2(\eta) =  a^2(\eta) \Omega^2 \l(\frac{k}{a(\eta)}\r) 
- \frac{a''(\eta)}{a(\eta)}.
\label{eq:tim_dep_fre}
\eeq
The modified dispersion relations must satisfy the property 
$k_{\rm mod} \sim k$ for $k << k_{\rm Pl}$.  We have tabulated the three
regimes of interest and the dominant contribution of 
$\varpi^{\rm mod}_k$ in these regimes in Table \ref{ta:sum}. Using the
notation in Table I, the solution to Eq. (\ref{eq:nonlin_scalar}) in
regime I, in the WKB limit is
\br
\chi^{I}_k(\eta)& =& \frac{A_k}{\sqrt{2 W^I_k(\eta)}} 
\exp\l( -i \int W^I_k(\eta) d\eta\r) \nonumber \\
&+& \frac{B_k}{\sqrt{2 W^I_k(\eta)}} \exp\l( i \int W^I_k(\eta) d\eta\r),
\label{eq:regimeI_1}
\er
where, $A_k$ and $B_k$ are constants and are determined by the choice
of the initial conditions. Orthonormality of the modes gives,
\beq
|A_k|^2 - |B_k|^2 = 1.
\label{eq:AkBk_cond}
\eeq
Here again, we assume $A_k$ and $B_k$ to be arbitrary functions of $k$
satisfying the above condition. In regime II, solution to
Eq. (\ref{eq:nonlin_scalar}) is (Minkowski) plane wave solution, i. e.,
\beq
\chi^{II}_k(\eta) = \alpha_k \exp(-i k \eta) + \beta_k \exp( i k \eta).
\label{eq:regimeII_1}
\eeq
The junction conditions of the wave modes $\chi^I_k/\chi^{II}_k$
and its derivatives at $\eta = \eta_{\rm Pl}$ gives
\br
\label{eq:alpha_gen}
\alpha_k& = & A_k~ \vartheta_k~ \exp[-i I(\eta_{Pl}) + i k\eta_{Pl}],\\  
\label{eq:beta_gen}
\beta_k & = & B_k~ \vartheta_k~ \exp[i I(\eta_{Pl}) - i k\eta_{Pl}],
\er
where, 
\br
\label{eq:def_vart}
{\vartheta_k}&=&  \sqrt{\frac{W_k^I(\eta_{Pl})}{8 k^2}} + 
\frac{1}{\sqrt{8 W_k^I(\eta_{Pl})}}, \\ 
\label{eq:def_I} 
I(\eta)& =& \int_{\eta_i}^{\eta} W_k^I(\eta') d\eta' = \int_{\eta_i}^{\eta} 
a(\eta') \Omega[k_{ph}(\eta')] d\eta'.
\er
In regime III, the solution is
\beq
\chi^{III}_k(\eta) = C_k^{\rm mod} a(\eta),
\label{eq:regimeIII_1}
\eeq
where $C_k^{\rm mod}$ is a constant whose modulus square gives the
power spectrum of the density perturbations. Using the junction
conditions at $\eta = \eta_H$, we get
\beq
C_k^{\rm mod} = \frac{1}{a(\eta_H)}
\l[\alpha_k \exp(-i k \eta_H) + \beta_k \exp( i k \eta_H)\r].
\label{eq:spec_gen1}
\eeq
Rewriting $\alpha_k/\beta_k$ in-terms of $A_k$ and $B_k$, we get
\br
& &\!\!\!\!\!\!\!\!\!\!\!\!\!\!\!\!\!\!\!\!\!\!|C_k^{\rm mod}|^2 =  a^{-2}~ 
\vartheta_k^2~\Bigg[ |A_k|^2 + |B_k|^2 \nonumber \\
& + & {\mathcal R}( A_k B^*_k \exp[-2 i I(\eta_{Pl}) + 
2 i k (\eta_{Pl} - \eta_H)]) \Bigg] 
\label{eq:sepc_gen3} 
\er
Using the condition (\ref{eq:AkBk_cond}), we obtain
\br 
& &\!\!\!\!\!\!\!|C_k^{\rm mod}|^2 =  \frac{H^2 k^{-2}}{4 k_{\rm mod}(\eta_H)} 
\l(1 + \frac{k_{\rm mod}(\eta_H)}{k}\r)^2 
\Bigg[ \frac{1}{2} +  |B_k|^2  \nonumber \\
\label{eq:maineq}
&&\!\!\!\!\!\!\!\! + \sqrt{|B_k|^4 + |B_k|^2}\cos \l[-2 I(\eta_{Pl}) 
+ 2 k (\eta_{Pl} - \eta_H)\r]\Bigg]. 
\er
Using the property that $k_{\rm mod} \simeq k$ for $k << k_{\rm pl}$,
we get
\br 
k^3 |C_k^{\rm mod}|^2 & = & \frac{H^2}{2} + H^2 |B_k|^2 + 
H^2 |B_k| \sqrt{|B_k|^2 + 1} \nonumber \\
\label{eq:C_K_nonlin}
&\times& \cos \l[- 2  I(\eta_{Pl}) + 2 k (\eta_{Pl} - \eta_H)\r]. 
\er
The above result shows that the spectrum of fluctuations has a
scale-invariant term plus a non scale-invariant term which strongly
depends on the initial state ($B_k$) and the form of the dispersion
relation ($\Omega$). If we assume the initial state of the field
(Minimum energy state) to be
\br
\label{eq:ini_mini1}
\chi_k(\eta_i)_{\rm mod}& =& \sqrt{\frac{1}{2 a(\eta_i)}} 
\Omega^{-1/2}\l(\frac{k}{a(\eta_i)}\r), \\ 
\label{eq:ini_mini2}
\frac{1}{i}\chi_k'(\eta_i)_{\rm mod}& =& \pm \sqrt{\frac{a(\eta_i)}{2}}
\Omega^{1/2}\l(\frac{k}{a(\eta_i)}\r),
\er
we get $A_k = \exp[I(\eta_i)], B_k = 0$; which results in a flat
scale-invariant spectrum.

\subsection{Analysis and Interpretation of results}

In subsections (\ref{subsec:standard}) and (\ref{subsec:dispersive}),
we obtained the power spectrum of fluctuations for an {\it arbitrary}
initial state in standard inflationary scenario and dispersive field
theory models. In this subsection, we compare and interpret the
results obtained. We first show that the third term in the right hand
side of (\ref{eq:C_K_nonlin}) cancels out. We then show that the
signatures of the deviations from the flat scale-invariant spectrum
due to non-linear dispersion relations {\it cannot} be differentiated from
the standard inflationary scenario with an {\it arbitrary} initial state.

Using the property that the non-linear dispersion relations becomes
linear for $k_{\rm Pl} >> k_{\rm ph}$, we can write them in the
general form
\beq
\Omega = k_{\rm Pl} F\l(\frac{k_{\rm ph}}{k_{\rm Pl}}\r),
\eeq
where $F[k_{\rm ph}/k_{\rm Pl}] \to k_{\rm ph}/k_{\rm Pl}$ in the large
wavelength limit. Substituting the above form in (\ref{eq:def_I}),
we get
\beq
I(\eta_{\rm Pl}) = \frac{k_{\rm Pl}}{H} \times Q
\label{eq:I_main}
\eeq
where
\beq
Q = \int_{H \eta_i}^{H \eta_{\rm Pl}} \frac{d (H \eta)}{( H \eta)} 
F\l(\frac{k (H \eta)}{k_{\rm Pl}}\r)
\eeq
Using the typical energy scale at the start of inflation to be GUT
scale [$E \approx 10^{15} GeV$, $\eta_i \approx - 10^{-36}
sec$, $H\approx 10^{11} GeV^{-1}$], the prefactor $k_{\rm Pl}/H$ in
the expression (\ref{eq:I_main}) is of the order $10^{10}$.  This
implies that unless the integral $Q$ is a small number (of the order
$10^{-9}$), the argument of the $cosine$ term in (\ref{eq:C_K_nonlin})
rapidly oscillates and can be set to zero. [This is because, in the
finding $C_l$'s (spherical harmonics of the temperature anisotropy)
we need to integrate over $k$ and the $cosine$ term vanishes.]

Assuming the amount of inflation to be $80-e$ foldings, we get
$\eta_{\rm Pl} \approx - 10^{-38} sec$. To obtain a rough estimate of
the integral $Q$, we consider four dispersions relations
(\ref{eq:paddy_dis} -- \ref{eq:ahlu_dis})-- Padmanabhan's,
$\kappa$-Poincare, Unruh's and Ahluwalia's. The numerical value of the
integral $Q$ in all these cases ranges between $0.5$ to $0.7$. [For
numerical work, we use MATHEMATICA owing to its convenience and
flexibility it offers.] Since, the argument of $cosine$ in the
expression (\ref{eq:C_K_nonlin}) is large (of the order $10^{9}$) for
all practical purposes, it can be set to zero. Thus,
Eq. (\ref{eq:C_K_nonlin}) reduces to
\beq
k^3 |C_k^{\rm mod}|^2  =  \frac{H^2}{2} + H^2 |B_k|^2.
\label{eq:C_K_fin_non}
\eeq
The following features are noteworthy regarding this result:
\begin{enumerate}
\item If the field modes are expanded as WKB modes and the initial state 
of the field is $A_k = 1; B_k = 0$, then the spectrum of fluctuations
are flat scale-invariant for any general dispersion relation. [In the
case of Corley/Jacobson dispersion relation, $\Omega$ becomes
imaginary for $k > k_{\rm Pl}$ and hence the expansion of modes as WKB
modes is not valid.] In the standard inflationary scenario, the 
fluctuations are flat if the initial state of the field is 
$a_k = 1; b_k = 0$. 

\item For an arbitrary initial state, the spectrum of fluctuations in 
the dispersive field theory models have a scale-invariant term plus a
non-scale invariant term. The non-scale invariant term {\it strongly}
depends on the initial state ($B_k$). However, the power spectrum does
not depend {\it strongly} on the dispersion relation. Thus, the form
of the dispersion relation does not play significant role in obtaining
the corrections to the scale-invariant spectrum.

\item In our discussion, we have assumed $|b_k|$ and $|B_k|$ to be 
arbitrary functions of $k$. The form of $|b_k|$ and $|B_k|$ contribute
to the non scale-invariant part of the power spectrum. Setting,
\beq
|B_k|^2 = |b_k|^2 + |b_k| (|b_k|^2 + 1)^{1/2},
\eeq
we observe that the spectrum of fluctuations in the standard
inflationary scenario and in dispersive field theory model are
same. Hence, the signatures of the deviations from the flat
scale-invariant spectrum from the CMBR observations due to quantum
gravitational effects {\it cannot} be differentiated from the standard
inflationary scenario with an {\it arbitrary} initial state. In other
words, the signatures of the deviation from the flat scale-invariant
spectrum is {\it not necessarily} due to the trans-Planckian effects.
Even in the standard inflation, the deviations from the flat spectrum 
can be obtained for a different choice of vacuum state -- 
other than the Bunch-Davies vacuum.

\end{enumerate}

\section{Conclusions}

We have studied the quantum gravitational effects on the spectrum of
primordial density perturbation in the inflationary epoch. We have
incorporated the quantum gravitational corrections using non-linear
dispersion relations to the quantum scalar field i. e., the scalar
field satisfying the linear wave equation with higher spatial
derivative terms propagating in the de Sitter space-time [and ignored
the back-reaction of the particles on the metric]. We keep the form of
the dispersion relation to be arbitrary and have not assumed any
specific dispersive field theory model.

We have performed a general analysis of the power spectrum for a class
of dispersion relations whose field modes can be expanded as WKB modes
over a range of initial conditions for the quantum field. We have
shown that the spectrum of fluctuations due to the quantum
gravitational corrections has a scale-invariant term plus a non
scale-invariant term which strongly depends on $B_k$ (the initial
state) and the form of the dispersion relation $\Omega$. However, the
strong dependence on the dispersion relation cancels out due to the
rapidly varying phase term. We have thus shown that the form of the
dispersion relation do not play significant role in obtaining the
corrections to the scale invariant spectrum. We have also shown that
the deviations from the flat spectrum due to quantum gravitational
effects (\ref{eq:C_K_nonlin}) {\it cannot} be differentiated as opposed to the
deviations from the flat spectrum for an {\it arbitrary initial state} in
the standard inflationary scenario (\ref{eq:C_K_lin}).

\vspace{-1.22cm}

\begin{acknowledgments}
The author would wish to thank T. Padmanabhan and K. Subramaniam for
useful discussions. The author is being supported by the University
Grants Commission, India through Senior Research Fellowship.
\end{acknowledgments}

\end{document}